\documentclass[
    ,final  ]
  {aipproc}

\layoutstyle{8x11double}

\begin{document}

\def\gtrsim{\mathrel{\hbox{\rlap{\hbox{\lower4pt\hbox{$\sim$}}}\hbox{$>$}}}}

\title{Pulsation Powered by Deuterium Burning in Brown Dwarfs and Very-Low-Mass 
Stars}

\classification{97.10.Sj,97.20.Vs,95.75.Wx}
\keywords{brown dwarfs,low-mass stars, star clusters, stellar pulsations, stellar 
photometry, light curves}

\author{Ann Marie Cody}{
  address={California Institute of Technology, MC 249-17, Pasadena, CA 91125}
}

\begin{abstract}
Pulsation powered by deuterium burning in brown dwarfs and very 
low mass stars has been put forth (Palla \& Baraffe 2005) as a novel probe of 
the interiors of these objects in the 1-15 Myr age range. Previous 
observations have hinted at variability on the expected timescales of a few 
hours, suggesting but not confirming that the phenomenon is at work in young 
brown dwarfs. We have recently carried out a dedicated campaign to 
search for this putative class of pulsators among known low-mass members of 
five young star clusters. Our survey achieves sensitivity to periodic 
oscillations with photometric amplitudes down to several millimagnitudes. 
We present the census of variability over timescales ranging from minutes to 
days and discuss the current prospects for pulsation as a tool in the study of 
young, objects near the substellar boundary. As a byproduct, this work provides
new insights into the distribution of stellar rotation periods at young ages via 
the detection of variability due to cool surface spots. 
\end{abstract}

\maketitle


\section{Pulsation as a Window into Young Brown Dwarfs}

With masses between approximately 0.013 and 0.08 M$_{\odot}$ (13.6-83.8 M$_{\rm Jup}$), brown 
dwarfs (BDs) provide a crucial link between stellar and planetary evolution. Theoretical 
models \cite{1997ApJ...491..856B,2003A&A...402..701B} are moderately successful in 
reproducing the observed properties of BDs, at least at intermediate and old ages 
\cite{2001RvMP...73..719B}. Such models can be used to estimate fundamental parameters, 
namely mass and age.  However the models have not been extensively tested by independent 
measurements of these fundamental quantities, nor do they account in detail for more complex 
realities such as magnetic fields and rotation which can have significant effects as 
illustrated for very low mass stars by \cite{2000ApJ...543L..77D}. Over the past dozen years, 
substantial populations of brown dwarfs have been identified at relatively young ages 
\cite{1996ApJ...469L..53R} and in star forming regions (e.g., \cite{1997ApJ...489L.165L}; 
\cite{2000ApJ...541..977N}; \cite{2004ApJ...610.1045S}). They are observed to host intriguing 
phenomena such as disk accretion and mass outflow, similar to their more massive brethren, in 
addition to magnetic activity and atmospheric weather, common at all ages among BDs.  What is 
needed now is correlation of theoretical predictions with observational phenomena in order to 
better constrain -- using fundamental physics -- the basic parameters of brown dwarfs.  We 
discuss here one such possible new method.

Brown dwarfs are not typical targets for stellar seismology. However, the wealth of recently 
discovered young cluster populations as well as new theoretical modeling efforts offer incentive 
to search for pulsations in these objects. Palla \& Baraffe (\cite{2005A&A...432L..57P}; 
hereafter PB05) have proposed that young ($\sim$1--15~Myr) BDs and very low mass stars (VLMSs) 
are subject to radial instability as a result of central deuterium burning and its strong 
temperature dependence (e.g., the $\epsilon$ mechanism). Based on non-adiabatic models, they 
predict significant mode growth in a mass range from the BDs (M$\sim$0.02~M$_\odot$) to the 
VLMSs (M$\sim$0.1~M$_\odot$). The corresponding pulsation periods are expected to lie between 
one and five hours. Using a different set of models from \cite{1997ApJ...491..856B} and a linear 
pulsation code, we have independently verified these basic results. Even more encouraging is the 
substantial overlap between available observational data for young BDs and VLMSs and the 
pulsation instability strip proposed by PB05, presented in Fig.\ 1.

Although theoretical predictions provided the initial motivation to search for pulsation in 
young, low-mass cluster members, some observational evidence of periodic variability has been 
reported on the appropriate $\sim$hour timescales. Several particularly promising pulsation 
candidates have been uncovered in the $\sigma$~Orionis cluster \cite{1999A&A...348..800B}, 
\cite{2003A&A...408..663Z}, and \cite{2004A&A...424..857C}. Such variability is unlikely to 
be the result of rotational modulation of surface spots, since the short periods are 
inconsistent with rotational velocities previously derived for a handful of these objects 
\cite{2005ApJ...626..498M}. 

We present here a dedicated observational campaign to photometrically monitor previously 
identified low-mass cluster members. Our observations are aimed at confirming the 
phenomenon of pulsation in young BDs, and could lead to the first anchoring of the low-mass end 
of interior models - upon which all studies of the initial mass function and star formation 
history are based.

\clearpage

\section{Observational Campaign}
\subsection{Search Space}

We have set out to perform a comprehensive photometric survey of BDs and VLMSs in a number of 
1-5~Myr star clusters. At these ages, objects below 0.1~$M_\odot$ experience the deuterium 
burning stage of the pre-main sequence, making them potential pulsation candidates. We assembled 
a target sample from a set of confirmed and likely members of the IC~348 cluster, 
Upper Scorpius, Taurus, Chamaeleon~I, and the $\sigma$~Orionis cluster. Spectral types are 
available for many objects, and we have employed the calibration of \cite{2003ApJ...593.1093L} 
to convert them to the temperature scale underpinning the brown dwarf models in PB05 
\cite{2003A&A...402..701B}. With a combination of photometric and spectroscopic data, we 
are able to place most known very low-mass young cluster members on the H-R diagram. The result, 
as shown in Fig.\ 1, indicates a promising number of BDs and VLMSs on or near the instability 
strip. Because of measurement errors and possible model-dependent systematic effects in the 
position of the strip, we have sought to survey a large swath of parameter space above the 
zero-age main sequence.

Not only must our survey target objects favorable for pulsation, but its design must also take 
into account expected variability properties and limits.  The linear theory of pulsation in BDs 
and VLMSs predicts periods of $\sim$1--4 hours, but cannot determine photometric variability 
amplitudes or the effect of convection on pulsation. We therefore designed observations to probe 
for pulsation to several millimagnitudes on timescales of a few hours. Since 
sensitivity to periodic signals depends on both signal-to-noise (S/N) and total number of data 
points, the detection limits reached with a 1-meter class telescope observing over one or two 
weeks are roughly equivalent to those achieved with a few nights with a larger telescope. We 
have selected the former set-up, since the longer time baseline offers increased frequency 
resolution and lower probability of weather interruptions.

The majority of our observations were carried out with the CTIO 1.0-meter telescope and 
the robotic Palomar 60-inch telescope. For each selected cluster field, we observe for a 
total duration of 10-20 nights with continuous monitoring at 5-15 minute cadence (apart 
from occasional weather interruptions). We have aimed for 1\% or better photometric 
precision, enabling identification of periodic variability in BDs with amplitudes as low 
as 0.005 magnitudes. Data are primarily in the $I$ band, where late-type objects are 
relatively bright but light is not dominated by potential dust disks. 

These observations provide an unprecedented combination of cadence, precision, and 
comprehensive coverage of very low mass cluster members. As such, we have set out to confirm 
or place limits on the existence of pulsation by covering a statistically significant sample of 
deuterium-burning objects across different masses and ages. The campaign is nearing 
completion, with observations thus far of most BDs and VLMSs in IC~348, $\sigma$ Orionis, and 
Upper Scorpius, as well as about half of the intended survey regions in Cha I.
Table~1 lists details on the five young clusters we have targeted, 
as well as observational time acquired; numbers of BDs and VLMSs listed refer to the {\em 
entire} cluster, and we list the fraction of these monitored.

\begin{table}
\begin{tabular}{cccccc}
\hline
   \tablehead{1}{c}{b}{Cluster\\Region}&
   \tablehead{1}{c}{b}{Approx.\\Age}&
   \tablehead{1}{c}{b}{Known BDs \\ \& VLMSs}&
   \tablehead{1}{c}{b}{Telescopes}&
   \tablehead{1}{c}{b}{Nights\\Observed}&
   \tablehead{1}{c}{b}{Known BDs\\ covered}  \\
\hline
IC 348 & 2-3~Myr & 32 & Palomar 60'' & 10 & 75\%\\
                 &         &    &  & & \\
$\sigma$ Orionis & 3-5~Myr & 29 & CTIO 1m, & 27 & 75\%\\
                 &         &    & Palomar 200'' & & \\       
Cha I & 1-3~Myr & 11 & CTIO 1m  & 13 & 25\%\\
                 &         &    &  & & \\
Upper & 5~Myr & 5 & Palomar 60'', & 20 & 2\%\\
Scorpius           &       &   & CTIO 1m       &  & \\
                 &         &    &  & & \\
Taurus & 1-3~Myr & 2 & Palomar 60''  & 12 & 2\% \\
\hline
\end{tabular}
\caption{Survey Coverage}
\label{tab:a}
\end{table}

\begin{figure}
  \includegraphics[height=.32\textheight]{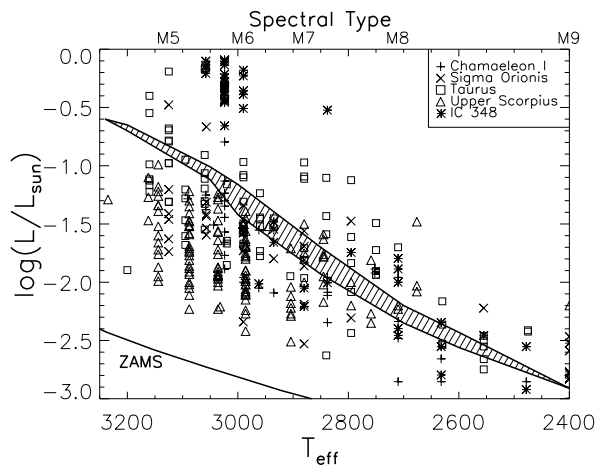}
  \caption{Many young cluster members are now known to fall on or near the 
deuterium-burning pulsation instability strip from PB05 (hashed area). Plotted here are 
effective temperatures and luminosities (in units of $L_{\odot}$) for BDs and VLMSs in 
the five clusters included in this photometric survey. As expected from the $\sim$1--5 Myr 
age estimates, all points are well above the zero age main sequence (ZAMS). Data were drawn 
primarily from \cite{2003ApJ...593.1093L}, \cite{2007ApJS..173..104L}, 
\cite{2003A&A...404..171B}, \cite{2000AJ....120..479A}, \cite{2006ApJ...645..676L}, and 
\cite{2006AJ....132.2665S} and have typical errors of 100~K in temperature, and 0.15~dex in 
log luminosity.}
\end{figure}

\subsection{Variability Detection}
We have produced lightcurves for data obtained thus far with a combination of aperture and 
image-subtraction photometry. In non-crowded fields, we employ a modified version of the 
variable-aperture photometry program {\em Vaphot} \cite{2001phot.work...85D}. Initial 
aperture sizes are chosen on an object-by-object basis so as to maximize the S/N ratio. 
Aperture sizes are then scaled with the seeing full-width-half-maximum (FWHM) in subsequent 
images. This approach conserves the fraction of stellar flux measured while maintaining 
optimal S/N. For objects with close companions whose light contaminates the aperture or sky 
background, we have instead relied on the {\em Hotpants} image subtraction package 
\cite{2004ApJ...611..418B} to remove flux from all non-variable objects. We then perform 
aperture photometry on the output residual images. These procedures have resulted in 
photometric performance very close to the limits imposed by poisson noise and sky background. 
The typical lightcurve RMS as a function of $I$-band magnitude is presented in Fig.\ 2. The 
resulting light curves display several different types of variable behavior which we have 
investigated for signs of pulsation.

According to the theory of PB05, pulsation in BDs and VLMSs is expected to 
induce brightness fluctuations at a single frequency, which in theory can be identified with 
a power spectrum. However, it is not known a priori how many of the objects 
in our sample are host to multiple types of variability, some of which may be 
high-amplitude and non-periodic. The plot of object magnitude versus lightcurve RMS in Fig.\ 2 
offers a quick visual assessment of the global variability properties of our BDs and VLMSs. 
Objects whose lightcurves are inconsistent with a constant source appear as outliers, and we  
can statistically select them using a $\chi^2$ test. We model the median measurement 
uncertainty as a function of magnitude with the estimated poisson noise, sky background, 
and a small systematic error component. For variability detection at >99\% certainty, 
lightcurves must have a $\chi^2$ value greater than 6.6.  Equivalently, we identify 
all data points lying more than a factor of 2.58 above the median photometric uncertainty 
as very likely variables.

To probe lightcurves for lower-level variability and search specifically for periodic 
signals, we employed the program Period04 \cite{2005CoAst.146...53L}, which computes an 
oversampled discrete fourier transform and may also be applied to time series 
with gaps. Periodic lightcurve variations appear as peaks in this periodogram. Since 
variability is concentrated in a narrow range of frequency space, we are able to probe to 
amplitudes at or below the level of the photometric precision, particularly at periods of <5 
hours where noise is mostly uncorrelated. Based on \cite{1993A&A...271..482B}, we have 
selected a criterion of S/N>4 in the periodogram (where signal is the amplitude of 
variability) for detection of periodic variables at 99\% confidence. For typical BDs with 
$I\sim17$ and an expected pulsation period of 1--4 hours, we are sensitive to variability 
amplitudes down to several millimagnitudes, as seen from the periodogram noise level in Fig.\ 4.

\begin{figure}
  \includegraphics[height=.25\textheight]{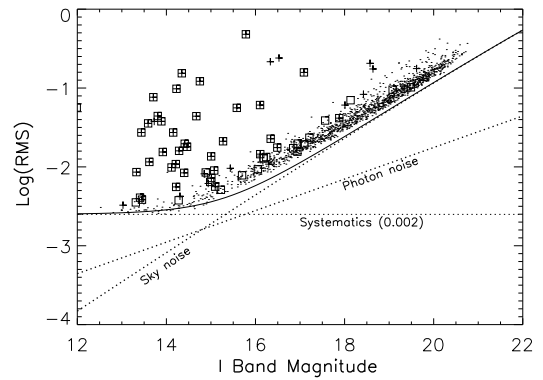}
  \caption{Example of photometric performance as a function of 
magnitude, from our data on the $\sigma$~Orionis cluster. Dots represent photometry of the 
entire field, while boxes denote likely cluster members and crosses mark objects that have 
passed one or more of our variability detection criteria.}
\end{figure}

\section{Results}
\subsection{Rotation and Accretion-related Variability}
While this study was initially designed to search for the putative new effect of pulsation in 
deuterium-burning objects, the large dataset has also uncovered numerous other phenomena, such 
as brightness modulation by rotating stellar spots, accretion-related fluctuations, and 
eclipses by companions. These effects are present in objects across a wider range in mass than 
is expected for pulsation. Thus we can analyze and classify variable behavior in slightly 
higher mass stars ($M\sim0.1-0.5 M_{\odot}$), and use the results to help interpret 
lightcurves obtained for the subsample of BDs and VLMSs.

We find that within the M0-M9 spectral type range (e.g., $M\sim0.02-0.5 M_{\odot}$), fully 50\% 
of young low-mass cluster members surveyed are variable, as evident in Fig.\ 2. Approximately 
one third of these exhibit variations at the $\sim$10\% level but lack detectable 
periodicities. As shown in the example at the top of Fig.\ 3, such aperiodic objects are likely 
accretors. In contrast, most of the remaining variables are unambiguously periodic with 
amplitudes at the few percent level and likely indicative of rotating spots, as seen at the 
bottom of Fig.\ 3. Their period distribution peaks near 1.8 days, with a steep drop-off at 15 
hours and a tail toward the observing duration of 10--15 days. The average amplitude of 
periodic variables is 0.03 magnitudes, although this value may be biased by our sensitivity 
limits.

\begin{figure}
  \includegraphics[height=.5\textheight]{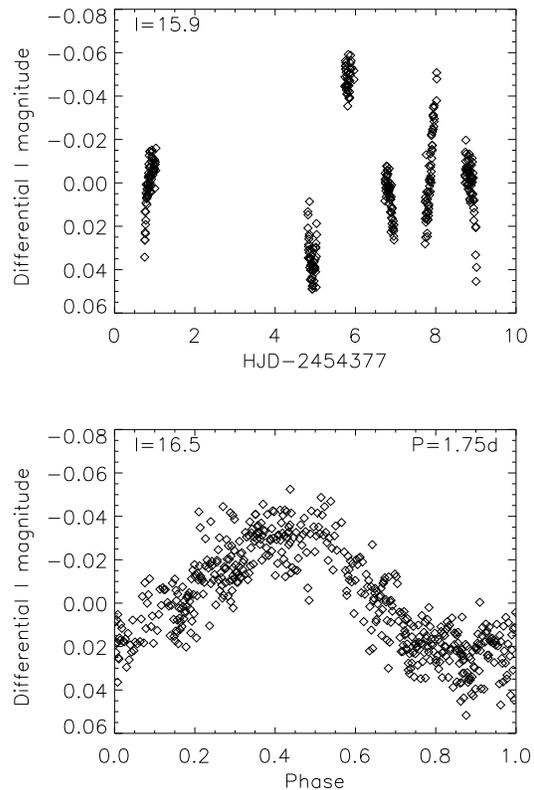}
  \caption{Lightcurves for two very-low-mass stars in IC~348 and 
$\sigma$ Orionis. {\em Top:} High-amplitude variability in this VLMS was confirmed with 
the $\chi^2$ test, whereas power spectrum analysis did not reveal any periodic 
frequencies. The stochastic nature of the variability thus suggests that the object is 
actively accreting. {\em Bottom:} A typical periodic variable, with 0.03-magnitude amplitude. 
The lightcurve fluctuates with a period of 42 hours, a timescale suggestive of rotating spot 
features.} 
\end{figure}

\subsection{Signs of Pulsation in Brown Dwarfs?} Periodic variability at the few percent level 
is certainly common amongst low-mass pre main sequence cluster members. But can any of it be 
attributed to pulsation instead of spots? In the absence of multi-band data, {\em timescale} is 
the main factor available to distinguish the two phenomena. While rotation periods of less than 
one day are not unheard of for young very low mass stars, variability on 1--2-hour timescales 
is exceedingly short to be explained by the passage of rotating spots. Typical rotation periods 
seen amongst young cluster objects seem to fall around 1-10 days \cite{2007prpl.conf..297H}, 
although there is some evidence that this decreases into the brown dwarf regime 
\cite{2009arXiv0906.2419R}. In addition, rotational velocities corresponding to a 1-hour period 
would reach 100 km~s$^{-1}$ and higher, uncomfortably close to the break-up speeds estimated 
for objects at these ages. Consequently, we consider any objects with variability periods under 
$\sim$5~hours to be prime pulsation candidates.

Despite sensitivity to millimagnitude amplitudes, the results of our periodogram analysis 
indicate a dearth of variables with periods less than 0.7 day. Furthermore, most of the brown 
dwarf targets show a surprising lack of variability on {\em any} timescale.  Contrary to 
previous reports (\cite{1999A&A...348..800B}; \cite{2003A&A...408..663Z}; 
\cite{2004A&A...424..857C}), we do not observe definitive variability with periods shorter than 
15 hours. This is somewhat unexpected given the copious spot-related variability observed in 
slightly higher mass objects, and may be indicative of physical trends in the surface 
properties of brown dwarfs. 

Interpretation of these results depends crucially on how many objects in the sample are actually 
on the narrow instability strip. Because of field-of-view constraints, we have not had the 
luxury of specifically targeting many of the best pulsation candidates. But observational 
uncertainties (typically $\sim$0.15 dex in log luminosity and 100~K in temperature), possible 
intracluster age spreads, and the uncertain location of the strip itself also create difficulty 
in determining {\em which} objects are the best pulsation targets. Assuming that the width of 
the instability strip is approximately correct as calculated by PB05, we estimate that less than 
30\% of the BDs and VLMSs in our photometric sample falls on or near it. If we count only 
objects that are directly {\em on} the strip, this fraction drops to 22\%. Out of $\sim$50 BDs 
and VLMSs in the sample, we therefore night expect $\sim$10--15 pulsators.

In reality, we have identified just a handful of cases thus far in which possible periodicities 
appear in the periodogram at 1--5-hour timescales. All of these are just above the detection 
limit at S/N$\gtrsim$4.0; an example is presented in Fig.\ 4. Pulsation may be present in these 
4--5 objects, but it is at such a low level that confirmation with a different instrument 
set-up is required. Future observational plans include follow-up observations of some of these 
most promising targets with both the Spitzer Space Telescope and Hubble Space Telescope.

In conclusion, we have presented an extensive photometric monitoring campaign on 1--5~Myr BDs 
and VLMSs, in hopes of confirming whether some of these objects are host to a new class of 
pulsation. While current evidence is not definitive, upcoming space-based data may provide the 
final word on variability in young, very-low-mass objects. 

\begin{figure}
  \includegraphics[height=.5\textheight]{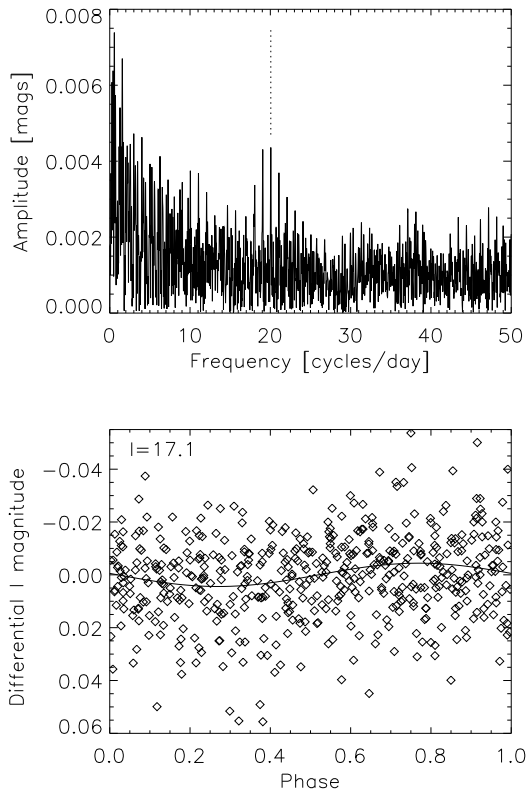}
  \caption{{\em Top}: The periodogram for a young brown dwarf in our sample shows a potential
periodicity (marked with the dashed line) with amplitude 0.0044 magnitudes at 20.1
cycles/day (1.2 hours). At low frequencies the data suffer from correlated (``red'') noise,
but beyond 10 cycles day$^{-1}$ the noise level in the periodogram reaches 0.001 magnitudes.
Hence this tentative detection appears at signal-to-noise of just over 4.0. {\em Bottom}:
The light curve is phased to the possible signal. Due to the marginal nature of the
detection, the signal cannot be confirmed by eye.}
\end{figure}



\bibliographystyle{aipproc}   

\vspace{4cm}

\bibliography{pulsation_cody}

\IfFileExists{\jobname.bbl}{}
 {\typeout{}
  \typeout{******************************************}
  \typeout{** Please run "bibtex \jobname" to optain}
  \typeout{** the bibliography and then re-run LaTeX}
  \typeout{** twice to fix the references!}
  \typeout{******************************************}
  \typeout{}
 }

\end{document}